\documentclass[twocolumn,showpacs,aps,prd,superscriptaddress]{revtex4}

\usepackage[none]{hyphenat}

\usepackage{amsmath}
\usepackage{amssymb}
\RequirePackage{xspace}
\usepackage[utf8]{inputenc}
\usepackage[caption=false]{subfig}

\usepackage[dvipsnames]{xcolor}

\usepackage{tikz}
\usetikzlibrary{positioning,arrows}
\usetikzlibrary{decorations.pathmorphing}
\usetikzlibrary{decorations.markings}

\usepackage{pgfplots}
\pgfplotsset{compat=newest}
\usepackage{xparse}
\usetikzlibrary{positioning,arrows,patterns}
\usetikzlibrary{decorations.markings}
\usetikzlibrary{calc}

\newcommand{\als}{\ensuremath{\alpha_s}\xspace}
\newcommand{\ale}{\ensuremath{\alpha_e}\xspace}

\newcommand{\cOale}{\ensuremath{{\cal O}(\ale)}\xspace}

\newcommand{\dq}{\ensuremath{d_q}\xspace}
\newcommand{\dtildeq}{\ensuremath{\widetilde{d}_q}\xspace}
\newcommand{\dc}{\ensuremath{d_c}\xspace}
\newcommand{\dtildec}{\ensuremath{\widetilde{d}_c}\xspace}
\newcommand{\db}{\ensuremath{d_b}\xspace}
\newcommand{\dtildeb}{\ensuremath{\widetilde{d}_b}\xspace}

\newcommand{\CP}{\ensuremath{\mathcal{CP}}\xspace}

\newcommand{\ecm}{e\,\text{cm}}
\newcommand{\cm}{\text{cm}}

\begin{document}

\title{Improved bounds on heavy quark electric dipole moments}

\author{H.~Gisbert}
\email{hector.gisbert@tu-dortmund.de} 
\author{J.~Ruiz Vidal}
\email{Joan.Ruiz@ific.uv.es} 
\affiliation{IFIC, Universitat de Val\`encia-CSIC, Valencia, Spain}

\date{June 9, 2020}
\begin{abstract}
    New bounds on the electric dipole moment (EDM) of charm and bottom quarks are derived using the stringent limits on their chromo-EDMs. The new limits, 
    $|\dc|<1.5\times10^{-21}\:\ecm$ and $|\db|< 1.2\times 10^{-20}\:\ecm$,
    improve the previous ones by about three orders of magnitude. These indirect bounds have implications for different models of new physics, including two-Higgs-doublet, leptoquarks, and supersymmetry models.
    
\end{abstract}

  
\maketitle

Searches for electric dipole moments (EDMs) are currently setting stringent constraints on models of new physics (NP) with additional \CP-violation sources~\cite{Chupp:2017rkp,Yamanaka:2017mef,Pospelov:2005pr,Engel:2013lsa,Dekens:2014jka}. Since the standard model predictions are well below the current experimental accuracy, any signal of a non-zero EDM would be a clear sign of NP. 
Moreover, the persisting B-anomalies suggest a non-trivial flavor structure in NP models, which can enhance the heavy quark EDMs~\cite{Buttazzo:2017ixm,Dekens:2018bci}.
Due to their very small lifetime, direct EDM searches on heavy-flavoured hadrons represent an experimental challenge and only indirect limits on heavy quark dipole couplings have been obtained to date.
However, this situation may change with the new proposals to search for the EDM of charmed and bottom baryons at the LHC~\cite{Botella:2016ksl,Bagli:2017foe,Bezshyyko:2017var,Baryshevsky:2016cul}.
In this Letter, a new approach for setting indirect bounds on quark EDM couplings is presented. By exploiting the mixing of operators under the renormalization group and using current constraints on the chromo-EDM of charm and bottom quarks~\cite{Sala:2013osa,Chang:1990jv}, we extract new bounds on their corresponding EDMs that improve the current ones by several orders of magnitude. 

For that purpose, let us consider the following flavour-conserving \CP-violating effective Lagrangian
\begin{align}
    \mathcal{L}_{\text{eff}}\:=\:\sum_{i=1}^2\sum_{q}C_i^q(\mu)\:O_i^q(\mu)\:+\:C_3(\mu)\:O_3(\mu)~,
\end{align}
where the index $q$ runs over the relevant flavours at the chosen renormalization scale. The effective operators are defined as
\begin{align}
O_1^q\:&\equiv\:-\:\frac{i}{2}\:e\:Q_q\:m_q\:\bar{q}^\alpha\:\sigma^{\mu\nu}\gamma_5\:q^\alpha\:F_{\mu\nu}~,\nonumber\\
O_2^q\:&\equiv\:-\:\frac{i}{2}\:g_s\:m_q\:\bar{q}^\alpha\:\sigma^{\mu\nu}\:T_a\:\gamma_5\:q^\alpha\:G^a_{\mu\nu}~,\\
O_3\:&\equiv\:-\:\frac{1}{6}\:g_s\:f_{a b c}\:\epsilon^{\mu\nu\lambda\sigma}\:G^{a}_{\mu\rho}\:G^{b \rho}_{\nu}\:G^{c}_{\lambda\sigma}\nonumber~,
\end{align}
where $Q_q$ and $m_q$ are the quark charge and quark mass, respectively.
The quark EDM, chromo-EDM, and the usually defined coefficient $\omega(\mu)$ of the Weinberg operator are related to the Wilson coefficients by
\begin{align}
\dq(\mu)\:&=\:e\:Q_q\:m_q(\mu)\:C_1^q(\mu) ~,\nonumber\\
\dtildeq(\mu)\:&=\:m_q(\mu)\:C_2^q(\mu)~,\\
\omega(\mu)\:&=\:-\:\frac{1}{2}\:g_s(\mu)\:C_3(\mu)~.\nonumber
\end{align}

When a heavy quark is integrated out, its chromo-EDM gives a finite contribution to the Weinberg operator~\cite{Braaten:1990gq,Chang:1990jv,Boyd:1990bx}, which is strongly constrained from the limits on the neutron EDM.
This allows to bound the quark chromo-EDMs to be~\cite{Sala:2013osa,Chang:1990jv},
\begin{align}
|\dtildec(m_c)| &< \: 1.0 \times 10^{-22}\:\cm~, \nonumber\\
|\dtildeb(m_b)| &< \: 1.1 \times 10^{-21}\:\cm~. \label{eq:bounddtildeq}
\end{align}
Attempts to constraint heavy quark EDMs have followed different strategies:
flavor-mixing contributions into light quark EDMs~\cite{Sala:2013osa,CorderoCid:2007uc,Grozin:2009jq},
$b\to s \gamma$ transitions~\cite{Sala:2013osa}, 
mixing into the electron EDM via light-by-light scattering diagrams~\cite{Grozin:2009jq} and
tree-level contributions to the ${e^+e^-\to q\:\bar{q}}$ total cross section~\cite{Escribano:1993xr,Blinov:2008mu}. These approaches yield results between $10^{-13}$ and $10^{-17}\:\ecm$, the most restrictive ones being~\cite{Sala:2013osa,Blinov:2008mu}
\begin{align}
|\dc(m_c)| &< \: 4.4 \times 10^{-17}\:\ecm~, \nonumber\\
|\db(m_b)| &< \: 2.0 \times 10^{-17}\:\ecm~. \label{eq:prevbounddq}
\end{align}
In this work we follow a new strategy that relates the EDM and chromo-EDM operators in order to find new limits on \dq from the already available strong bounds on \dtildeq. This relation is done in a model-independent way using the renormalization group equations, which mix the effective operators when the energy scale is changed. 
The relevant diagrams include photon loops which have been neglected in previous works due to its small size compared with pure QCD corrections. Nevertheless, they represent the first non-zero contribution to the mixing we are interested in.\\

\begin{figure}[t]
    \centering
        \raisebox{-.5\height}{
			\includegraphics[width=0.5\columnwidth]{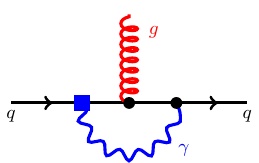}
        }
        \caption{The quark EDM coupling (blue square) induces a chromo-EDM through photon-loop diagrams. These represent the leading contribution to the matrix element $(\gamma_e)_{12}^{(0)}$.}
        \label{fig:feynmandiagram}
\end{figure}

The evolution of the Wilson coefficients is given by
\begin{align}
\frac{\text{d}}{\text{d}\ln\mu}\:\overrightarrow{C}(\mu)\:=\:\widehat{\gamma}^{\text{T}}\:\overrightarrow{C}(\mu)~,\label{eq:RGE}
\end{align}
where $\overrightarrow{C}\equiv(C_1^q,\:C_2^q,\:C_3)$ and $\widehat{\gamma}$ is the anomalous dimension matrix. This matrix can be expanded in powers of the QCD and QED coupling constants, $\als$ and $\ale$, respectively,
\begin{align}
\widehat{\gamma}\:=\:\frac{\als}{4\:\pi}\:\gamma_s^{(0)}\:+\:\left(\frac{\als}{4\:\pi}\right)^2\:\gamma_s^{(1)}\:+\:\frac{\ale}{4\:\pi}\:\gamma_e^{(0)}\:+\:\cdots~,
\end{align}
where $\gamma_s^{(0)}$ and $\gamma_s^{(1)}$ represent the one- and two-loop QCD corrections, while $\gamma_e^{(0)}$ encodes the one-loop QED correction~\cite{Weinberg:1989dx,Wilczek:1976ry,Braaten:1990gq,Degrassi:2005zd,Jenkins:2017dyc}. At $\mathcal{O}(\als^2)$, the quark EDM does not mix into the chromo-EDM and the first contribution only appears at $\mathcal{O}(\ale)$ from photon-loop diagrams as shown in Figure~\ref{fig:feynmandiagram}. Applying the standard techniques for the computation of anomalous dimensions~\cite{Buchalla:1995vs,Buras:1998raa} we obtain the matrix element $(\gamma_e)_{12}^{(0)}=8$, in agreement with the recent calculation in~\cite{Jenkins:2017dyc}.

Solving Eq.~\eqref{eq:RGE} by adding this contribution, the evolution of the charm and bottom chromo-EDMs read
\begin{align}
    \dtildec(m_c)\:&=\:-\:0.04\:\frac{\dc(M_{\text{NP}})}{e}\:+\:0.74\:\dtildec(M_{\text{NP}})~,\label{eq:cCEDMmixing}  \\
    \dtildeb(m_b)\:&=\:0.08\:\frac{\db(M_{\text{NP}})}{e}\:+\:0.88\:\dtildeb(M_{\text{NP}})~,\label{eq:bCEDMmixing}  
\end{align}
where we have taken $M_{\text{NP}}\sim 1\:$TeV as the scale of NP.
In this result, we have neglected the mixing of the Weinberg operator into the chromo-EDM due to the very strong bounds on $\omega$ from constraints on the neutron EDM~\cite{Pospelov:2005pr,Baker:2006ts}.
The mixing of \dtildeq into itself, described by the second term on the right-hand side of Eqs.~\eqref{eq:cCEDMmixing} and \eqref{eq:bCEDMmixing}, has leading contributions from pure QCD corrections, then corrections of \cOale can be safely neglected.

\begin{figure}[ht]
    \centering
    \includegraphics[width=0.8\linewidth]{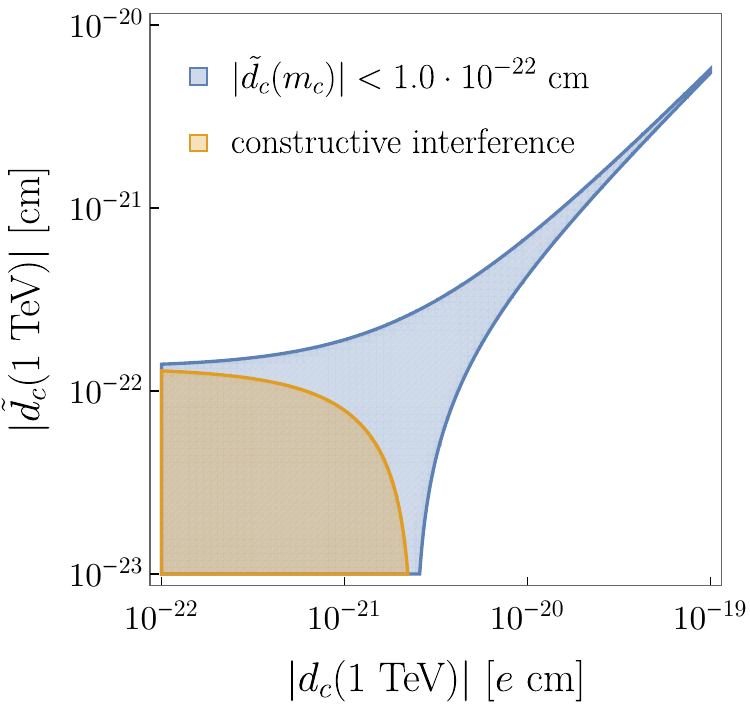}\\
    \vspace{0.3cm}
    \includegraphics[width=0.8\linewidth]{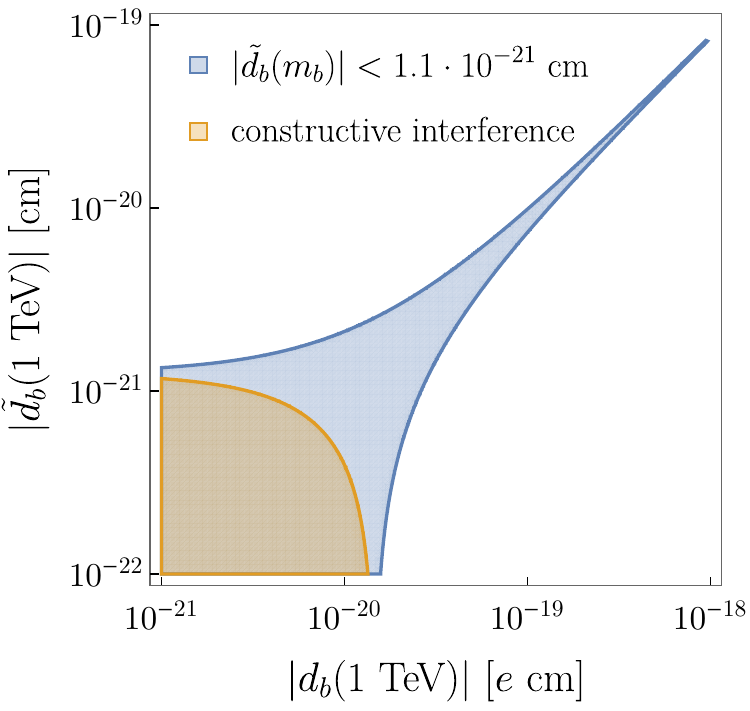}
    \caption{Bounds on the charm (bottom) chromo-EDM constrain the \dtildec-\dc (\dtildeb-\db) plane to the allowed blue area. The narrow region results from strong cancellation effects that are not present in the case of constructive interference, displayed in orange.}
    \label{fig:cEDMvsEDM}   
\end{figure}

Using the bounds on the chromo-EDMs at the low scales quoted in Eq.~\eqref{eq:bounddtildeq}, the parameter space on the \dtildeq-\dq plane is constrained as shown in Figure \ref{fig:cEDMvsEDM}. Strong fine-tuned cancellations between the two pieces of Eqs.~\eqref{eq:cCEDMmixing} and \eqref{eq:bCEDMmixing} result in an allowed region extending along a straight line which is unlikely to be realised in NP models.

Hence, we assume constructive interference between the EDM and chromo-EDM contributions at the NP scale to extract bounds on $\dq(M_{\text{NP}})$. Then, using the evolution of the EDM operator to bring these bounds down to the quark mass scale, the new bounds on the charm and bottom quark EDMs are
\begin{align}
|\dc(m_c)| &< \: 1.5\times 10^{-21}\:\ecm~,\nonumber\\
|\db(m_b)| &< \: 1.2 \times 10^{-20}\:\ecm~,
\label{eq:newlimits}
\end{align}
which improve the previous ones quoted in Eq.~\eqref{eq:prevbounddq} by three and four orders of magnitude, respectively.
This approach does not improve the current bounds on the top quark EDM~\cite{Cirigliano:2016njn,Fuyuto:2017xup} given that the limit on its chromo-EDM is of similar size~\cite{Kamenik:2011dk}.
These results directly depend on the chromo-EDM bounds, in Eq.~\eqref{eq:bounddtildeq}, which are obtained from the neutron EDM by neglecting cancellations between the light quarks (C)EDM and the Weinberg operator. The large uncertainty on the Weinberg operator contribution to the neutron EDM is treated conservatively by taking the smallest value within the confidence interval. 
We should point out that using the mercury EDM provides better bounds by about a factor 2~\cite{Graner:2016ses,Engel:2013lsa}. However, given the additional sources of uncertainty together with the cancellation effects that may arise between the several contributions to the mercury EDM, we consider only the direct experimental bounds on the neutron EDM. 
Note also that higher values of the NP scale yield less conservative results, e.g. a 30\% stronger bounds for $M_{\text{NP}} =10 \:\text{TeV}$.
The inclusion of dimension-six four-quark operators would add extra terms in the right-hand side of Eqs.~\eqref{eq:cCEDMmixing} and \eqref{eq:bCEDMmixing}. The resulting cancellation effects are nevertheless smaller than the self-correction of the chromo-EDM, shown in Figure~\ref{fig:cEDMvsEDM}.

\begin{figure}[ht]
    \centering 
    \includegraphics[height=0.3\columnwidth]{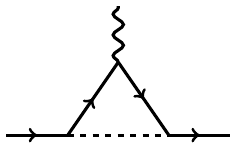} \hspace{0.3cm}    
    \includegraphics[height=0.3\columnwidth]{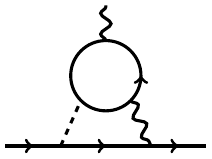}

	\caption{Diagrams with new scalars (dashed lines) contributing to the quark dipole moments. The external wavy lines represent photons (gluons) for the contribution to the quark EDM (chromo-EDM), while the internal one can be either a photon, gluon, or weak gauge bosons.}
	\label{fig:2HDMdiagrams}
\end{figure}

In the following we evaluate the effect of the new bounds for the charm and bottom quark EDMs on the parameter space of different theories beyond the standard model (BSM).

In the context of minimal flavour violation, the EDM of different quarks only differ by the quark mass. When this dependency goes as $d_q \propto m_q$, the strong bounds on the light quark EDMs, ${|d_{u,d}|\lesssim10^{-25}\,\ecm}$, impose stronger constraints than our bounds. However, if the quark EDM scales with larger powers of the quark mass, the heavy quark EDMs are greatly enhanced and become competitive.

This is the case of the Two-Higgs-Doublet model (2HDM), which generates fermion EDMs via the Yukawa couplings of new scalars. To avoid flavour-changing neutral-currents at tree level, which are very constrained at the TeV scale, we restrict the discussion to the 2HDM with flavour alignment~\cite{Pich:2009sp,Pich:2010ic,Penuelas:2017ikk} in which the Yukawa matrices $Y_{d,u}$ are proportional to the quark mass matrices,
\begin{align}
Y_{d}\,=\,\varsigma_d\,M_d~,\quad Y_{u}\,=\,\varsigma_u^\dagger\,M_u~,    
\end{align}
where $\varsigma_{u,d}$ are complex numbers and contain the \CP-violation.
In this type of models, the quark EDMs arise at one-loop level mediated by neutral ($S^0$) or charged scalars ($S^\pm$) (see Figure~\ref{fig:2HDMdiagrams}), giving contributions proportional to $m_{q}^3/M_{S^0}^2$ or $m_q m_{q'}^2/M_{S^{\pm}}^2 |V_{qq'}|^2$, respectively, where $V$ is the Cabibbo-Kobayashi-Maskawa matrix.
These mass factors suppress the light quark EDMs, which are actually dominated by two-loop Barr-Zee contributions as shown Figure~\ref{fig:2HDMdiagrams}. The EDM of heavy quarks are much larger and, even with weaker experimental bounds, they can be more restrictive.

\begin{figure}[t]
    \centering
    \includegraphics[width=0.7\linewidth]{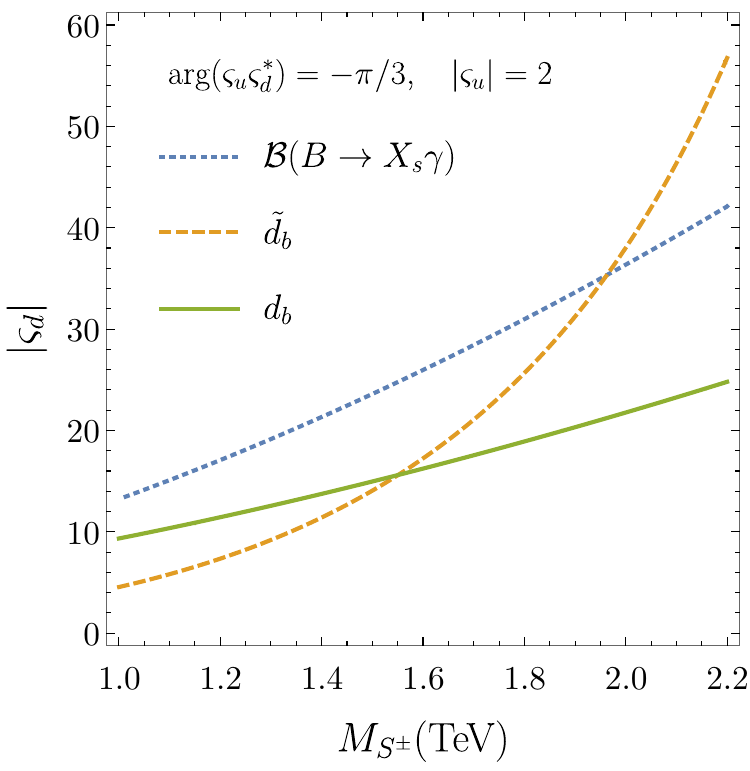}
    \caption{Constraints on the parameter space of the MW model. The allowed regions are below the lines. The lower-limit on the mass range follows from~\cite{Miralles:2019uzg}.}
    \label{fig:MW_zetaD-M}   
\end{figure}

Among these models, we shall consider the contribution to the bottom quark (chromo-)EDM by the colour-octet scalars appearing in the so-called Manohar-Wise (MW) model~\cite{Manohar:2006ga}. The relevant one-loop diagrams were computed in \cite{Martinez:2016fyd} and are dominated by the exchange of a charged scalar with mass $M_{S^\pm}$. 
The constraints on this model from the experimental results on the $B^0_s - \overline{B}^0_s$ mixing, $B_s^0\to \mu\mu$, or $B^0\to X_s\gamma$ were analyzed in~\cite{Cheng:2015lsa}. Among them, the inclusive branching ratio $\mathcal{B}(B\to X_s\gamma)$ dominates the constraints on the $|\varsigma_u\,\varsigma_d| - M_{S^\pm}$ plane. As it is shown in Figure~\ref{fig:MW_zetaD-M}, the bounds on the bottom EDM derived above are more restrictive than this observable and even surpass the constraining power of the chromo-EDM for $M_{S^\pm} \gtrsim 1.5 \text{~TeV}$.
This is always the case for the regions of parameter space with nonzero phases of \CP-violation, $\arg(\varsigma_u\varsigma_d^*)\gtrsim15^o$.

In recent years, a series of measurements hinting at the violation of lepton flavour universality have motivated new physics extensions with non-universal couplings between the three families. 
When the new \CP violation sources are specific to the quark family, the EDM of each quark carries independent and complementary information that should be used in complete phenomenological analyses of such models.

An example of family non-universal \CP-violating interaction is found in models with scalar leptoquarks. These models are currently receiving a lot of attention as they are able to explain naturally the deviations in $ b\to c\tau\bar{\nu_\tau}$ transitions
~\cite{Buttazzo:2017ixm,Murgui:2019czp,Becirevic:2016yqi,Cornella:2019hct,Fajfer:2012jt,Hiller:2016kry}.
The additional charged currents contributing to this process are parametrized through the coefficient $g_{S_L}$. Combining the experimental values of $R_D$ and $R_{D^*}$ results in allowed regions for $g_{S_L}$ away from the real axis~\cite{Becirevic:2018afm} which induce a sizeable charm EDM~\cite{Dekens:2018bci}.
If no signal is observed in the planned neutron EDM experiments with sensitivities
of few times $10^{-27} \ecm$~\cite{Chupp:2017rkp}, the resulting upper limits on the charm EDM (extracted with the method presented here) will rule out this model as an explanation for the B-anomalies.
In fact, the authors of~\cite{Dekens:2018bci} already draw exclusion regions due to the charm EDM. Their results are nevertheless based on lattice QCD calculations for the strange quark tensor charge, whose translation into the charm quark is highly uncertain.
The next BSM extension we discuss is the minimal supersymmetric standard model (MSSM). Among the large number of free parameters that it contains, there are many new sources of \CP violation. It is customary to restrict phenomenological analyses to just two sources: the trilinear couplings $A$, and the $\mu$-term (see definitions in \cite{Martin:1997ns}). Since the fermion EDMs appear at one-loop level, these parameters are strongly constrained by the neutron and electron EDMs~\cite{Pospelov:2005pr}. Nevertheless, in more general cases the $A$ coupling can be a $3\times3$ matrix whose elements are specific to the quark family. In particular, the charm quark EDM accesses the element $A_c$ predominantly via gluino loops \cite{Aydin:2002ie}. Updating the numerical analysis of~\cite{Aydin:2002ie} by taking into account the LHC restrictions on the masses~\cite{ATLAS:2019jvl}, we still find values of $\dc \sim 10^{-20}\ecm$ for scharm masses $M_{c1}$($M_{c2}$) of $1(2)$ TeV, gluino mass $m_{\tilde{g}}=1.6$ TeV, and  $\arg(A_c)=\pi/4$. These regions of the parameter space are therefore excluded and the new bounds should be included in more detailed analyses of this model.

Beyond the MSSM, there are new \CP-violating sources that can generate contributions to quark EDMs. 
An example of these is the MSSM with gauged baryon and lepton numbers (BLMSSM). Scaling the results of~\cite{Zhao:2016jcx} accounting for the top quark EDM bounds~\cite{Cirigliano:2016njn,Fuyuto:2017xup} we obtain values of $\dc$ reaching $10^{-19} \ecm$, \textit{i.e.} two orders of magnitude above the new upper limit in Eq.~\eqref{eq:newlimits}. As a consequence, the new heavy quark EDM bounds impose stringent constraints on the additional $\CP$-violating phases of the BLMSSM. During the publication process of this paper, a detailed analysis of the BLMSSM demonstrated the restrictive power of the new bounds on this model \cite{Yang:2019aao}. 
In the R-parity violating supersymmetry, the EDM of heavy fermions are the only EDM observables that directly access the bilinear combinations of the third quark generation $\text{Im}(\lambda_{i 3 3}{\lambda_{i 3 3}'^*})$, for $i=1,2$~\cite{Yamanaka:2014nba}. Nevertheless, the leading contribution appears in this model at two-loop level~\cite{Yamanaka:2014mda} and it is suppressed in comparison with other supersymmetric extensions. For this reason, the bottom EDM is not yet competitive with other observables when considering the effect of one coupling $\lambda_{ijj}$ at a time,  but it could be used in global analyses to restrict extended regions of fine-tuned cancellations.

In the literature there exist other models giving predictions on heavy quark EDMs at the level of our bound or higher. As exemplary, we found works based on Composite Higgs~\cite{Panico:2016ull} and 2HDM with non-universal extra dimensions~\cite{Iltan:2004xr}. We will not comment further on them.

In summary, we have presented a simple way to access the quark EDM through the corresponding chromo-EDM. The method relies on the inclusion of photon-loop corrections \cOale in the renormalization group equations, which are often overlooked due to its small size. Nevertheless, these corrections provide a new window to access effective operators which are otherwise unconstrained.
We derived new upper limits for the charm and bottom quark EDMs, and showed the potential of these operators to constraint the parameters of NP models. These limits will provide valuable input for detailed phenomenological analyses of BSM physics.

~\\
We want to express our gratitude to  W. Dekens, A. Delhom, M. Cerdà-Sevilla, F. Martínez-Vidal, N. Neri and A. Pich for useful discussions.
This work has been supported in part by the Spanish State Research Agency and ERDF funds from the EU Commission [Grants FPA2017-84445-P, FPA2017-85140-C3-3-P, FPA2015-68318-R and FPA2014-53631-C2-1-P], by Generalitat Valenciana [Grant Prometeo/2017-053 and Grant Prometeo/2014-049], and by the Spanish Centro de Excelencia Severo Ochoa Programme [Grant SEV-2014-0398].
The work of H.G. is supported by a FPI doctoral contract [BES-2015-073138], funded by the Spanish Ministry of Economy, Industry and Competitiveness.

\end{document}